\documentclass[a4paper,fleqn,usenatbib]{mnras}

\usepackage[T1]{fontenc}
\usepackage{ae,aecompl}

\usepackage{graphicx}	
\usepackage{amsmath}	
\usepackage{amssymb}	

\title[Disc structure and planet mass]{Constraining the masses of planets in protoplanetary discs from the presence or absence of vortices - Comparison with ALMA observations}

\author[Hallam \& Paardekooper]{
P. D. Hallam\thanks{E-mail: p.d.hallam@qmul.ac.uk},
S.-J. Paardekooper
\\
Astronomy Unit, School of Physics and Astronomy, Queen Mary University of London, E1 4NS, UK
}

\date{Accepted XXX. Received YYY; in original form ZZZ}

\pubyear{2019}

\begin{document}
\label{firstpage}
\pagerange{\pageref{firstpage}--\pageref{lastpage}}
\maketitle

\begin{abstract}
A massive planet in a protoplanetary disc will open a gap in the disc material. A steep gap edge can be hydrodynamically unstable, which results in the formation of vortices that can act as tracers for the presence of planets in observational results. However, in a viscous disc, the potential formation of these vortices is dependent on the timescale over which the massive planet accretes mass and with a sufficiently long timescale it is possible for no vortices to form. Hence, there is a connection between the presence of vortices and the growth timescale of the planet and it may therefore be possible to exclude a planetary interpretation of observed structure from the absence of vortices. We have investigated the effect of the  planet growth timescale on vortex formation for a range of planet masses and viscosities and have found an approximate relation between the planet mass, viscosity and planet growth timescale for which vortices are not formed within the disc. We then interpret these results in the light of recent observations. We have also found that planets do not need to be close to a Jupiter mass to form vortices in the disc if these discs have low viscosity, as these can be caused by planets as small as a few Neptune masses.

\end{abstract}

\begin{keywords}
hydrodynamics -- instabilities -- planet-disc interactions -- protoplanetary discs
\end{keywords}



\section{Introduction}
Planets form in protoplanetary discs via the accretion of dust and gas. These planets will excite density waves in the disc material that transport angular momentum from the planet and deposit it in the disc, exerting a torque on the disc material. If the deposition of torque is close enough to the planet \citep{Lin&Papaloizou1993, Bryden1999} and the resultant torque is stronger than the viscous diffusion of the disc material \citep{Lin&Papaloizou1979,Goldreich&Tremaine1980,Takeuchi1996,Crida2006} then a low density region, a gap, is formed around the orbital location of the planet. These two criteria are known as the thermal and viscous criteria for gap opening.

Observational results over the past few years \citep[e.g.][]{ALMApartnership2015,Andrews2016,vanderMarel2019} have provided high resolution images and information about the features that exist across the lifetime of protoplanetary discs. One important feature we can see from these images is the presence of dark bands in the dust emission of the disc. A clear example of this can be seen in images of the protoplanetary disc HL Tau \citep{ALMApartnership2015}. While we do not currently have a definitive answer as to what causes these, there is a possibility they could be planetary in origin, gaps formed by massive planets within the disc. Unfortunately, even assuming these dark bands are the result of planet-disc interaction, there is a degeneracy in the planet masses and disc viscosities that can adequately explain them \citep[e.g.][]{Mulders2013}. If both the disc viscosity and the planet mass are unknown, almost any gap can be explained by a planet. Hence, we would benefit greatly from a method that would allow us to rule out a planet of a given mass from having formed a gap.

When a planet opens a gap in the disc, there is the potential for the excitement of instabilities in the disc material. The steep edge of a gap carved by a planet is prone to hydrodynamic instabilities such as the Rossby wave instability. The non-linear evolution of this instability can result in the formation of vortices at the gap edge \citep{Lovelace1999,Li2000}.

Vortices are potential explanations for brightness asymmetries seen at the edge of a gap or cavity in observational results \citep{vanderMarel2013,Fukagawa2013,Perez2014,Marino2015,Kraus2017,Dong2018,Cazzoletti2018}. Numerical simulations show that indeed massive planets can excite the Rossby wave instability \citep{DeValBorro2006,DeValBorro2007,Fu2014}. However, vortices can also be formed by processes that do not involve a planet \citep[e.g.][]{Lovelace1999,Li2000,Klahr&Bodenheimer2003,Lyra&Klahr2011,Regaly2012,Raettig2013,Lyra2014,Bae2015}. Despite this, it is not unbelievable that the observed vortices in the discs were formed as a result of planets we cannot see. In many of the simulations in which these vortices occur the formation of the planet is not accounted for. In other words, the planet is already at its maximum mass before it begins to open a gap in the disc. As the strength of the deposited torque is dependent on the mass of the planet, the gap will open slower for a planet growing in mass than for a planet at maximum mass, giving the disc time to adjust viscously. It has been shown by \cite{Hammer2017} that over sufficient growth timescales it is possible for a massive planet to open a gap without exciting the Rossby wave instability and therefore without producing vortices. Hence, it may be possible to rule out a planet of a certain mass as the creator of the gap based on the presence or absence of vortices in the disc. Additionally, recent investigation into the shape of planet induced vortices have accounted for the timescale over which the planet grows and have found that vortices can become elongated \citep{Hammer2017}. As a result elongated vortices have their own specific observational features \citep{Hammer2019} that have potentially been observed \citep{Perez2014,Cazzoletti2018}. However, the majority of observational results do not show the features that imply elongated vortices, although this may be due to the resolution of the observations  

In this paper we are investigating the required planet growth timescales for which no vortices are formed in a protoplanetary disc. We study this for a range of masses and viscosities, with the aim to find an approximate relation between these parameters. This approximate relation could then be used to estimate limits for parameters of observed protoplanetary discs, or to determine whether or not it is likely that a planet exists in these discs. This acts to constrain the mass of a planet that can be formed in a disc of a certain age and viscosity, based on the presence or absence of vortices in the disc. Therefore, we aim for our result to aid in breaking the degeneracy that exists in explaining observed disc structure using planets. 

This paper is arranged as follows. In Section \ref{Sec:BasicEq}, we discuss the relevant equations solved to simulate disc evolution. In Section \ref{Sec:NumSet}, we discuss the code and numerical setup for our simulations. In Section \ref{Sec:Res}, we present our results. In Section \ref{Sec:Comp}, we compare our findings to prior observational work. In Section \ref{Sec:Disc} we discuss our results and any assumptions made in our simulations. Finally, in Section \ref{Sec:Conc}, we present our conclusions.

\section{Basic Equations}\label{Sec:BasicEq}
In this paper we perform a parameter study to find the growth timescales for which the Rossby wave instability is not excited by gap formation in a protoplanetary disc. We investigate a wide range of viscosities, mass ratios and growth timescales and so perform this study in two spatial dimensions, making the approximation of a locally isothermal disc. As the Hill sphere of the planet is larger than the scale height for gap opening, the disc appears two dimensional with respect to the planet. Hence, it is safe to assume that the approximation of two dimensions is sufficient to draw meaningful conclusions from.

\subsection{Two dimensional protoplanetary disc}

The continuity equation for the evolution of a protoplanetary disc's surface density, $\Sigma$, is given by

\begin{equation}
	\frac{\partial\Sigma}{\partial t} + \nabla\cdot (\Sigma\mathbf{v}) = 0
\end{equation}
where $\mathbf{v}$ is the velocity field. We simulate the evolution of a protoplanetary disc's surface density due to the presence of a planet by solving the two dimensional Navier-Stokes equation for the motion of the disc,

\begin{equation} \label{eq:Nav_Stokes}
	\Sigma\left(\frac{\partial\mathbf{v}}{\partial t} + \mathbf{v}\cdot\nabla\mathbf{v}\right) = -\nabla P -\nabla\cdot\mathbf{T} - \Sigma\nabla\Phi,
\end{equation}
where $\mathbf{T}$ is the Newtonian viscous stress tensor, $P$ is the pressure and $\Phi$ is the gravitational potential of the planet and star system. A cylindrical coordinate system is used, such that $\mathbf{v} = (v_R, R\Omega)$ where $v_R$ and $\Omega$ are the radial and angular velocities at a given radius. We assume the disc is locally isothermal, with an equation of state $P = c_s^2\Sigma$ where $c_s(R)$ is the sound speed at cylindrical radius $R$. We impose a constant aspect ratio $h = H/R$ where $H$ is the disc scale height. The sound speed in the disc is given by:

\begin{equation}
	c_s = h\sqrt{\frac{GM_*}{R}},
\end{equation}
where $M_*$ is the mass of the star. The planet is held on a fixed circular orbit at $R=R_0$, with angular velocity $\Omega = \Omega_0$ and orbital period $P_0$.

\subsection{Growing the planet mass}\label{SubSec:GrowP}
In our simulations it is important that the planet mass is not constant, it should grow with time over a number of orbits. This is in an attempt to eliminate the excitation of instabilities and formation of vortices that otherwise would be present if the planet is initialised at its maximum mass. Therefore we grow the planet mass over a timescale $t_{\textrm{G}}$ for $t \leq t_{\textrm{G}}$ using the following relation

\begin{equation}\label{Eq:GrowP}
	q(t) = q_f\frac{1}{2}\left(1 - \textrm{cos}\left(\pi\frac{t}{t_{\textrm{G}}}\right)\right)
\end{equation}
where $q(t) = M_p(t)/M_*$, $q_f = q(t_{\textrm{G}})$.

\subsection{Rossby wave instability}
The Rossby wave instability can occur due to a steep gradient in the radial profile of the angular velocity of the disc. In a disc with a gap opening planet, this location corresponds to the gap edge in the surface density of the disc. Using a long enough planet growth timescale, we expect the gap to open slower and hence the disc viscosity has more time to smooth out the steep gap edges. This will result in a smaller gradient in the velocity profile of the disc at this location.

For the Rossby wave instability to be excited within an adiabatic disc there must be a local maximum or minimum in the key function \citep{Lovelace1999},

\begin{equation} \label{eq:keyfunc}
	\mathcal{L}(R) = \mathcal{F}S^\frac{2}{\gamma}
\end{equation}
where $\mathcal{F}$ is approximately the inverse of the potential vorticity, $S = P/\Sigma^\gamma$ is the entropy and $\gamma$ is the ratio of specific heats. As this function clearly has surface density dependence, the opening of a gap in the disc will have an effect on whether the Rossby wave instability is excited and we would expect to see a local maximum or minimum occurring near a steep gap edge in the surface density of the disc. The presence of a maximum or minimum in this function is only one prerequisite for the excitation of the Rossby wave instability, it must also exceed a threshold value before the Rossby wave instability is excited \citep{Lovelace1999,Ono2016}. Hence, from this function alone we cannot predict the Rossby stability of a system. In addition, our simulations are not adiabatic, they are locally isothermal and as such are in the limit of $\gamma \rightarrow 1$. It is therefore difficult to predict whether the Rossby wave instability will be present a priori and we therefore determine from simulations whether the Rossby wave instability has led to the formation of vortices by calculating the potential vorticity across the disc. The potential vorticity is given by,

\begin{equation}
	\xi = \mathbf{\hat{z}}\cdot\frac{\nabla\times\mathbf{v}}{\Sigma}.
\end{equation}
A local minimum in the potential vorticity of the disc indicates the presence of a vortex, which is the non-linear outcome of the Rossby wave instability.

In addition to observing the two dimensional potential vorticity distribution, we also investigate the orbit-to-orbit difference in the potential vorticity of the disc. This method is helpful when instabilities are marginal and more difficult to determine from the two dimensional distribution. As the vortices have a different orbital velocity to that of the planet, we will see them move on an orbit-to-orbit basis, therefore if we see a sudden increase in the orbit-to-orbit maximum potential vorticity difference we can imply the presence of the Rossby wave instability. 

\section{Numerical Setup}\label{Sec:NumSet}
Our simulations are run in two dimensions using the FARGO3D code. This is a magnetohydrodynamic code designed to simulate disc evolution in one to three dimensions by solving the hydrodynamic equations of motion. FARGO3D is a good choice for investigating gap opening planets as it uses a C to CUDA translator to allow simulations to be run on Graphics Processing Units (GPUs). GPUs have limited memory, but decrease computational time significantly, allowing moderate resolution simulations to be run for extended periods of time. For more details see \cite{FARGO3D}.

Our simulations use a disc model that extends over a radial domain of $0.3 \leq R/R_0 \leq 6.0$ and an azimuthal domain of $-\pi \leq \phi \leq \pi$ with $285$ by $596$ cells respectively, using logarithmic cell spacing in the radial direction and the planet located at $R = R_0$. The initial surface density is $\Sigma_\textrm{int} = \Sigma_0 (R/R_0)^{-1/2}$ with $\Sigma_0 = 2.67\times 10^{-3}$. The indirect potential due to the planet is accounted for and the mass of the disc is arbitrary, as the planet does not feel the disc material. To achieve this, FARGO3D has been modified so that the indirect potential caused by the disc can be turned off separately. Reflecting boundary conditions were used with wave killing zones in the regions $0.3 \leq R/R_0 \leq 0.585$ and $5.43 \leq R/R_0 \leq 6.0$ similarly to \cite{DeValBorro2006}, meaning that excited waves are damped before reaching the edge of the simulation. The planet is held on a fixed circular orbit with no migration or disc self gravity accounted for. We use a constant kinematic viscosity $\nu$ and a fixed aspect ratio $h$. We vary both the final mass ratio, $q_f$ and $\nu$ between simulations.

We run our simulations for a number of orbits beyond the chosen planet growth timescale, $t_{\textrm{G}}$, in order to ensure the Rossby wave instability is not excited shortly after the planet has reached its maximum mass. This depends on the gap formation timescale and viscous timescale of the disc and as a result will vary depending on the plant mass and viscosity, however is often of the order of a few hundred orbits.   

\section{Results}\label{Sec:Res}

\subsection{Method}
We run our simulations for a range of mass ratios, $1.5\times 10^{-4}\leq q \leq 10^{-3}$ and disc viscosities $10^{-6} \leq \nu \leq 10^{-5}$. For each combination of planet mass and viscosity, we investigate the planet growth timescale for which we do not observe the formation of vortices in the disc. We use an aspect ratio $h = 0.05$ for the majority of our simulations. We focus on this one aspect ratio to reduce the size of the parameter space we investigate, in order to lower the computational time necessary. However, we do investigate changing the aspect ratio in Section \ref{SubSec:AspRat}. 

\begin{figure}
	\includegraphics[width=\columnwidth]{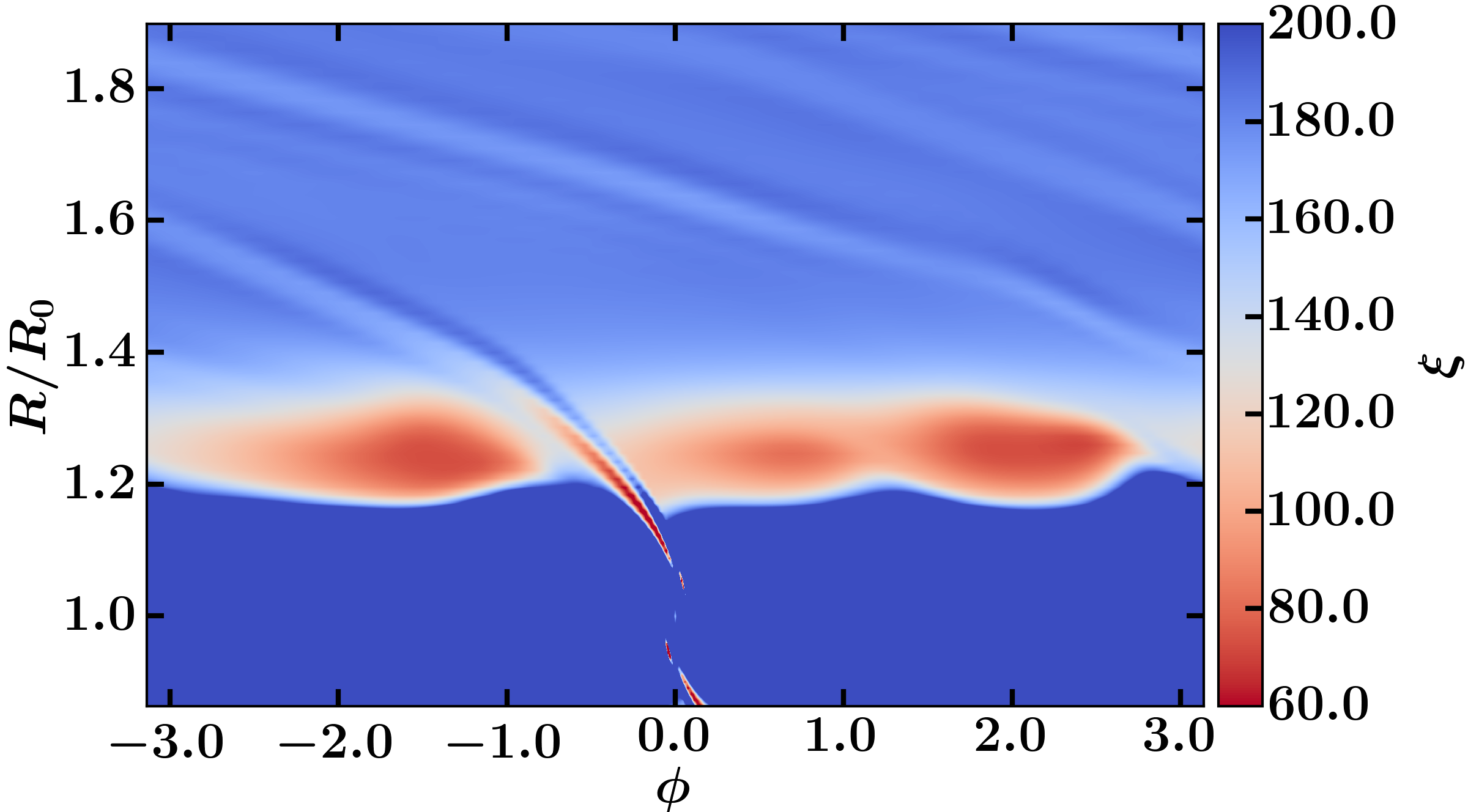}
	\vspace*{-5mm}
    \caption{An example of vortices clearly visible in the two dimensional potential vorticity distribution of the disc. In this case the disc has $\nu = 8\times 10^{-6}$ with a $q_f = 6\times 10^{-4}$ planet and a $t_{\textrm{G}} = 0$, meaning that the planet is initialised in the disc at $t = 0$ at its final mass.}
    \label{Fig:rwi}
\end{figure}

In order to determine the timescale at which vortices are not formed we calculate the potential vorticity across the disc. From this we determine the orbit-to-orbit maximum potential vorticity difference. As the vortices orbit at a different speed to the planet they move on an orbit-to-orbit basis. An extreme example of these vortices can be seen in Figure \ref{Fig:rwi}, which shows the formation of vortices in a $t_{\textrm{G}} = 0$ setup. This is an extreme case for the purpose of showing the vortices and using a growth timescale $t_{\textrm{G}} > 0$ will reduce the magnitude of these vortices. Hence approaching the values of $t_{\textrm{G}}$ that stabilise the disc it will become more and more difficult to identify the presence of vortices from these plots. 

\begin{figure}
	\includegraphics[width=\columnwidth]{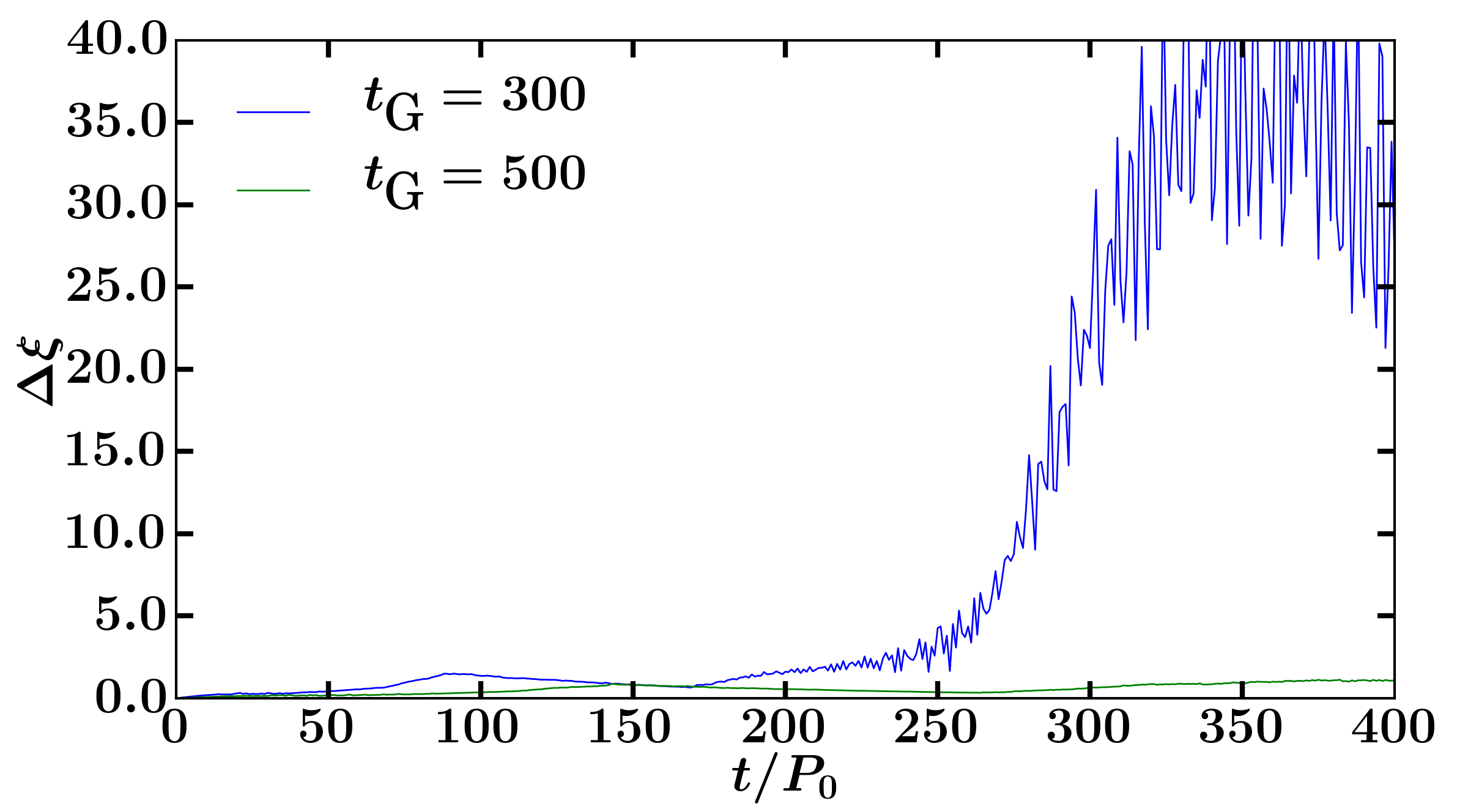}
	\vspace*{-5mm}
    \caption{Orbit to orbit maximum potential vorticity difference for a given growth timescale against the number of orbits elapsed. Illustrated here is the difference between a stable case, with $t_{\textrm{G}} = 500$ and a boundary case, with $t_{\textrm{G}} = 300$. The parameters for these two simulations are $\nu = 10^{-5}$ and $q_f = 9\times 10^{-4}$. For comparison, a $t_{\textrm{G}} = 0$ results in a $\Delta\xi \approx 400 - 1000$ at it's most unstable}
    \label{Fig:orb_to_orb}
\end{figure}

Such marginal cases are still important to distinguish, as it is known that vortices are efficient at trapping dust and this could enhance the visibility of a vortex observationally \citep{Barge&Sommeria1995}. If these vortices form a sharp increase in the orbit-to-orbit potential vorticity difference this should imply the excitation of the Rossby wave instability. An example of using the potential vorticity difference to highlight the presence of vortices can be seen in Figure \ref{Fig:orb_to_orb}. In this case we can see the comparison between a stable simulation with $t_G = 500$ and a simulation with $t_G = 300$. Here $t_G = 300$ is the boundary beyond which vortices form. In the stable case the potential vorticity difference has some small variations but is overall roughly constant, while in the boundary case there is a noticable increase in the potential vorticity difference. The appearance of vortices signals an even larger potential vorticity difference, and as the system becomes more unstable the magnitude of this increase in potential vorticity difference becomes larger. For example, the vortices in Figure \ref{Fig:rwi} produce a maximum $\Delta\xi \approx 200-300$. This highlights how useful this method is to identifying these vortices.

\begin{figure*}
	\includegraphics[width=\textwidth]{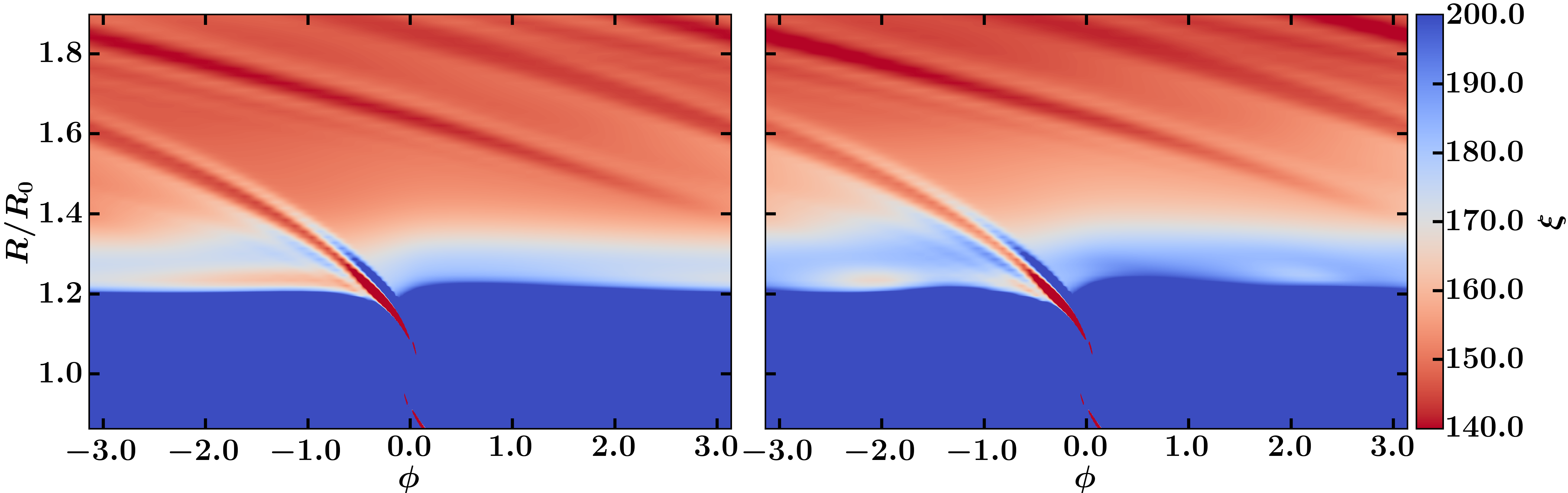}
	\vspace*{-5mm}
    \caption{A comparison showing the two dimensional potential vorticity distributions for a stable disc (left) and a disc containing vortices (right). Both of these discs show a $q_f = 9\times 10^{-4}$ planet in a $\nu = 8\times 10^{-6}$ disc at $t = 1500$. The difference between them is the stable case has a $t_{\textrm{G}} = 1400$ while the unstable case has $t_{\textrm{G}} = 1000$ and hence has excited an instability, which can be clearly seen from the presence of the vortex.} 
    \label{Fig:compare}
\end{figure*}

In particularly marginal cases, in which it becomes difficult to determine stability solely from the potential vorticity difference, we also use the two dimensional potential vorticity distribution to aid in determining whether or not the system is stable. An example of using the two dimensional potential vorticity distribution to determine if a vortex is present can be seen in Figure \ref{Fig:compare}, which shows two simulations with $t_G = 1400$ and $t_G = 1000$ both at $t = 1500$. While this is a considerably clearer case of vortex formation, this highlights the difference between a stable and unstable system.

We limit our potential vorticity difference calculations to the region close to the gap edge where the vortices are formed, excluding a region $\Delta\phi \approx 0.95$ either side of the planet. We do this because even in cases where there is little to no vortices small fluctuations in the location of the gap edge or the planet's wake, on the order of one to two radial cells can cause very large potential vorticity differences for steep gap edges. This can cause a false result implying the presence of vortices when there are none, hence we must take care to exclude this from our sampling region.

In addition, in some cases we find that weak features can be present in the potential vorticity distribution of the disc, even in simulations that are Rossby stable. These may be due to the interaction between orbiting material and the planet's outer wake, in which material orbiting past the planet's wake receives a decrease in potential vorticity flowing past the shock. This can be seen in the stable system in the left panel of Figure \ref{Fig:compare}. This mechanism was first identified by \cite{Koller2003}. These weak features often have no impact on the potential vorticity difference, even in the rare cases in which they stray out of the excluded zones.

\subsection{Error Estimates}
As we discretely sample the growth timescale, a higher sampling resolution will provide a more accurate value of the growth timescale at which vortices do not appear. Therefore, we choose errors on our selected growth timescale based on our sampling resolution. This should account for the possibility that the exact value we are looking for lies between our sampled growth timescales. In selecting whether a growth timescale results in the formation of vortices or not, we first make an educated guess at the growth timescale that results in no vortices and perform a simulation. We analyse the result of this, and increase or decrease the growth timescale by an amount, usually on the order of a few hundred orbits, but heavily dependent on the accuracy of the initial guess and the computational time required to complete the simulation. This is an iterative process, until we find the smallest growth timescale for which the system has no vortices. Hence, we have a discrete resolution in growth timescale, which is necessary to reduce computational time. The error due to this gets larger as the growth timescale increases, as these are more computationally expensive. Based on these considerations, for the purposes of this study we select errors as presented in Table \ref{Table:Errors}.

\begin{table}
\caption{Errors for different ranges of $t_{\textrm{G}}$.}
\centering
\begin{tabular}{c|c}
\hline
$t_{\textrm{G}}$ & $\Delta t_{\textrm{G}}$ \\
\hline
$0 \leq t_{\textrm{G}} \leq 500$& $50$ \\
$500 < t_{\textrm{G}} \leq 1000$& $100$ \\
$1000 < t_{\textrm{G}} \leq 2000$& $200$ \\
\hline
\end{tabular}
\label{Table:Errors}
\end{table}

\subsection{Minimum planet growth timescale}
\begin{figure*}
	\includegraphics[width=\textwidth]{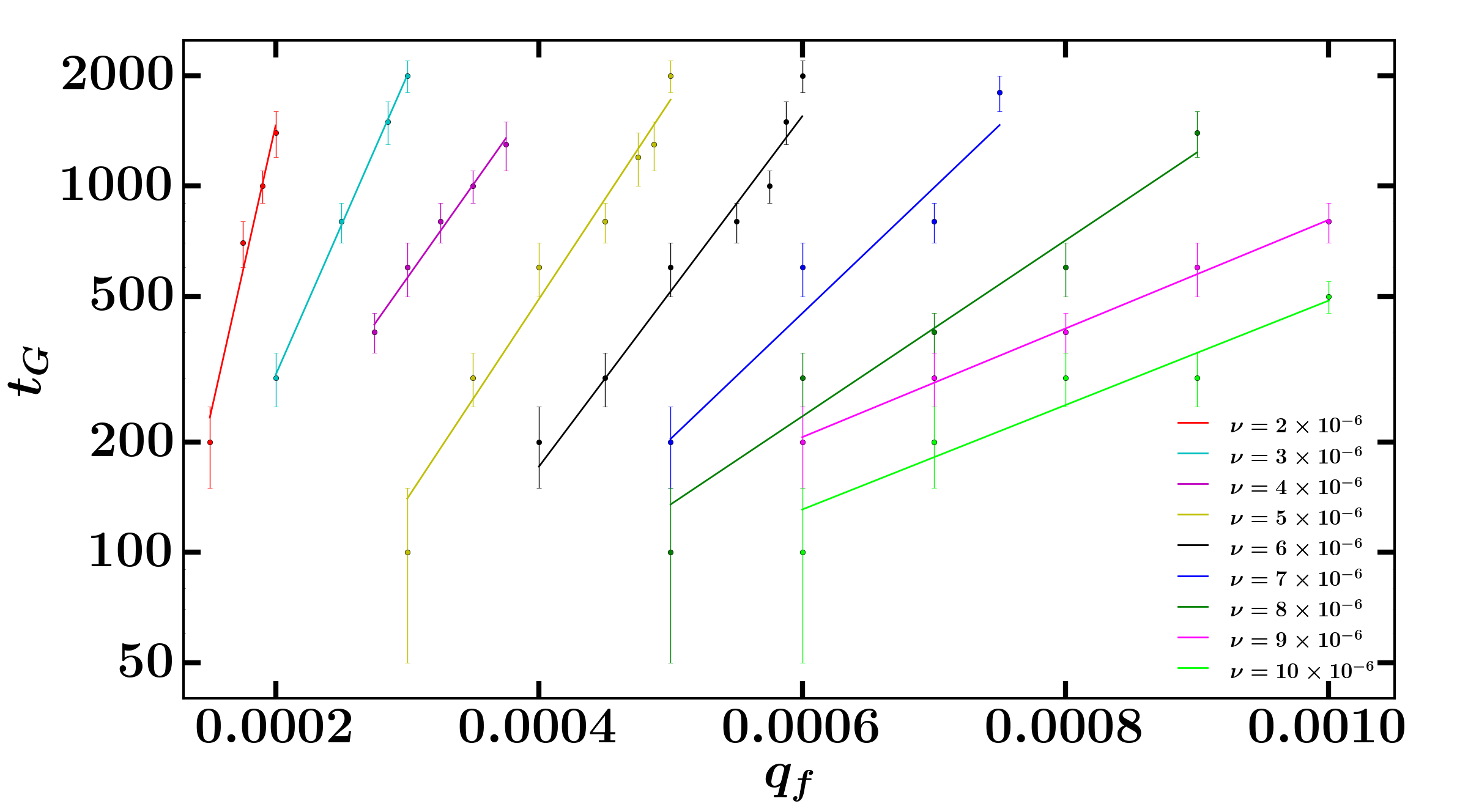}
	\vspace*{-5mm}
    \caption{Exponential best fit curves for the data presented in Table \ref{Table:pfTime} for each value of $\nu$, including errors decided upon as discussed in Section \ref{Sec:Res}, presented using a logarithmic scale in orbits. This clearly shows the trend of decreasing gradient with increasing viscosity.} 
    \label{Fig:log}
\end{figure*}

\begin{figure}
	\includegraphics[width=\columnwidth]{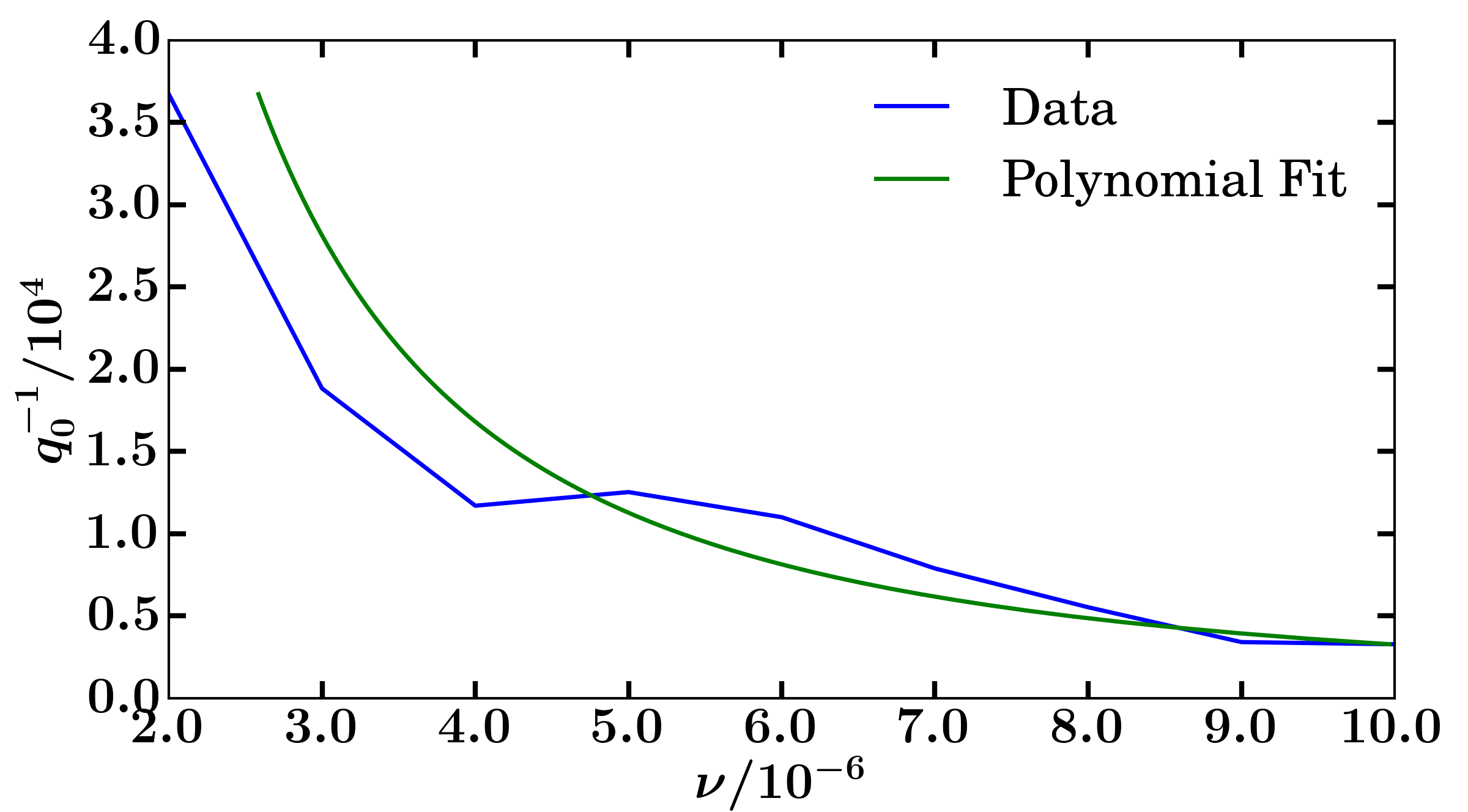}
	\vspace*{-5mm}
    \caption{Polynomial best fit curve for the gradients of the curves in Figure \ref{Fig:log} given by Equation \ref{Eq:Fit}. This shows relatively good agreement between the fit and the data in the mid to high viscosity range, however this gets worse at low viscosities and begins to significantly overestimate the gradient of the mass/growth timescale curve.} 
    \label{Fig:polyfit}
\end{figure}

Table \ref{Table:pfTime} shows the minimum planet growth timescales for which vortices are not formed, over the range of mass ratios and viscosities. This information is also presented in Figure \ref{Fig:log} which fits exponential curves of the form:

\begin{equation}\label{Eq:Pred_Fit}
	t = t_0e^{\frac{q_f}{q_0}}
\end{equation}
where $t_0$ and $q_0$ are constants dependent on the fit, given in Table \ref{Table:Cm}. From this figure we can see that the best fit exponential curves are a reasonable fit to the data and that as we increase the viscosity the curves become less steep and $q_0$ increases. From this we find a relation between $q_0$ and the viscosity, 

\begin{equation}\label{Eq:Fit}
	\nu = 9.1884\times 10^{-4} q_0^{0.5588}
\end{equation} 
This relation can be seen in Figure \ref{Fig:polyfit}. This shows a reasonable fit to the data in the mid to high viscosity regime, however deviation begins to become significant in the low viscosity regions. 

\begin{table}
\caption{Planet growth timescales for which no vortices are formed for different mass ratios and disc viscosities.}
\centering
\begin{tabular}{c|c|c}
\hline
Disc viscosity, $\nu$ &  Mass ratio, $q_f$ & Planet growth timescale, $t_{\textrm{G}}$ \\
\hline
$1.0 \times 10^{-6}$& $1.500 \times 10^{-4}$& $2000$ \\

$2.0 \times 10^{-6}$& $1.500 \times 10^{-4}$ & $200$\\
$2.0 \times 10^{-6}$& $1.750 \times 10^{-4}$ & $700$\\
$2.0 \times 10^{-6}$& $1.900 \times 10^{-4}$ & $1000$\\
$2.0 \times 10^{-6}$& $2.000 \times 10^{-4}$ & $1400$\\

$3.0 \times 10^{-6}$& $2.000 \times 10^{-4}$ & $300$\\
$3.0 \times 10^{-6}$& $2.500 \times 10^{-4}$ & $800$\\
$3.0 \times 10^{-6}$& $2.850 \times 10^{-4}$ & $1500$\\
$3.0 \times 10^{-6}$& $3.000 \times 10^{-4}$ & $2000$\\

$4.0 \times 10^{-6}$& $2.750 \times 10^{-4}$ & $400$\\
$4.0 \times 10^{-6}$& $3.000 \times 10^{-4}$ & $600$\\
$4.0 \times 10^{-6}$& $3.250 \times 10^{-4}$ & $800$\\
$4.0 \times 10^{-6}$& $3.500 \times 10^{-4}$ & $1000$\\
$4.0 \times 10^{-6}$& $3.750 \times 10^{-4}$ & $1300$\\

$5.0 \times 10^{-6}$& $3.000 \times 10^{-4}$ & $100$\\
$5.0 \times 10^{-6}$& $3.500 \times 10^{-4}$ & $300$\\
$5.0 \times 10^{-6}$& $4.000 \times 10^{-4}$ & $600$\\
$5.0 \times 10^{-6}$& $4.500 \times 10^{-4}$ & $800$\\
$5.0 \times 10^{-6}$& $4.750 \times 10^{-4}$ & $1200$\\
$5.0 \times 10^{-6}$& $4.875 \times 10^{-4}$ & $1300$\\
$5.0 \times 10^{-6}$& $5.000 \times 10^{-4}$ & $2000$\\

$6.0 \times 10^{-6}$& $4.000 \times 10^{-4}$ & $200$\\
$6.0 \times 10^{-6}$& $4.500 \times 10^{-4}$ & $300$\\
$6.0 \times 10^{-6}$& $5.000 \times 10^{-4}$ & $600$\\
$6.0 \times 10^{-6}$& $5.500 \times 10^{-4}$ & $800$\\
$6.0 \times 10^{-6}$& $5.750 \times 10^{-4}$ & $1000$\\
$6.0 \times 10^{-6}$& $5.875 \times 10^{-4}$ & $1500$\\
$6.0 \times 10^{-6}$& $6.000 \times 10^{-4}$ & $2000$\\

$7.0 \times 10^{-6}$& $5.000 \times 10^{-4}$ & $200$\\
$7.0 \times 10^{-6}$& $6.000 \times 10^{-4}$ & $600$\\
$7.0 \times 10^{-6}$& $7.000 \times 10^{-4}$ & $800$\\
$7.0 \times 10^{-6}$& $7.500 \times 10^{-4}$ & $1800$\\

$8.0 \times 10^{-6}$& $5.000 \times 10^{-4}$ & $100$\\
$8.0 \times 10^{-6}$& $6.000 \times 10^{-4}$ & $300$\\
$8.0 \times 10^{-6}$& $7.000 \times 10^{-4}$ & $400$\\
$8.0 \times 10^{-6}$& $8.000 \times 10^{-4}$ & $600$\\
$8.0 \times 10^{-6}$& $9.000 \times 10^{-4}$ & $1400$\\

$9.0 \times 10^{-6}$& $6.000 \times 10^{-4}$ & $200$\\
$9.0 \times 10^{-6}$& $7.000 \times 10^{-4}$ & $300$\\
$9.0 \times 10^{-6}$& $8.000 \times 10^{-4}$ & $400$\\
$9.0 \times 10^{-6}$& $9.000 \times 10^{-4}$ & $600$\\
$9.0 \times 10^{-6}$& $1.000 \times 10^{-3}$ & $800$\\

$1.0 \times 10^{-5}$& $6.000 \times 10^{-4}$ & $100$\\
$1.0 \times 10^{-5}$& $7.000 \times 10^{-4}$ & $200$\\
$1.0 \times 10^{-5}$& $8.000 \times 10^{-4}$ & $300$\\
$1.0 \times 10^{-5}$& $9.000 \times 10^{-4}$ & $300$\\
$1.0 \times 10^{-5}$& $1.000 \times 10^{-3}$ & $500$\\
\hline
\end{tabular}
\label{Table:pfTime}
\end{table}

\begin{table}
\caption{Values of the constants $t_0$ and $q_0$ for different disc viscosities.}
\centering
\begin{tabular}{c|c|c}
\hline
Disc viscosity, $\nu$ & $t_0$ & $q_0$ \\
\hline

$2.0 \times 10^{-6}$& $0.9469$ & $2.7238\times 10^{-5}$ \\
$3.0 \times 10^{-6}$& $7.0582$ & $5.3100\times 10^{-5}$ \\
$4.0 \times 10^{-6}$& $16.7741$ & $8.5449\times 10^{-5}$ \\
$5.0 \times 10^{-6}$& $3.2684$ & $7.9798\times 10^{-5}$ \\
$6.0 \times 10^{-6}$& $2.1016$ & $9.0870\times 10^{-5}$ \\
$7.0 \times 10^{-6}$& $3.9502$ & $1.2675\times 10^{-4}$ \\
$8.0 \times 10^{-6}$& $8.5080$ & $1.8079\times 10^{-4}$ \\
$9.0 \times 10^{-6}$& $26.5984$ & $2.9289\times 10^{-4}$ \\
$1.0 \times 10^{-5}$& $18.2717$ & $3.0466\times 10^{-4}$ \\
\hline
\end{tabular}
\label{Table:Cm}
\end{table}

As $t_0$ does not follow a simple law (see Table \ref{Table:Cm}) we cannot predict how $t_0$ evolves with viscosity. Hence, we cannot determine a continuous relation between the viscosity and planet growth timescale. This, in combination with Equation \ref{Eq:Fit}'s poor fitting at low viscosities means that we use Equation \ref{Eq:Pred_Fit} at our discrete viscosity intervals to form a relation between mass and viscosity. We can then use Equation \ref{Eq:Pred_Fit} evaluated at these viscosities in order to make statements about the potential masses of gap opening planets and viscosities of the disc based on the age of the system in question, using observational results regarding the detection or absence of vortices within these discs.

From Figure \ref{Fig:log} we can see that for higher viscosity discs it is possible to form more massive planets over shorter timescales, without creating vortices. Consider the planets formed with $q_{f} = 6\times 10^{-4}$. At $\nu = 10^{-5}$, it is possible to form this planet in only $100$ orbits with no vortices created. However, as we lower the viscosity the growth timescale to form this planet without creating vortices increases. Hence, at $\nu = 6\times 10^{-6}$ the same planet takes $2000$ orbits to form without creating vortices.

Another important result we can see in Figure \ref{Fig:log} pertains to the masses of planets that can form vortices in the disc. The exponential best fit curves in Figure \ref{Fig:log} show the minimum growth timescale required for a planet of a given mass to form without creating vortices. As we can see there is large parameter space at lower viscosities in which planets can create vortices while forming. Indeed, it is not even necessary for such planets to be particularly massive. For example consider a disc with $\nu = 2\times 10^{-6}$. In this disc, if a planet of $q_f = 2\times 10^{-4}$ is to form over a timescale $t_{\textrm{G}} < 1400$, it will create vortices in the disc. This planet is a similar mass to Saturn, and significantly less massive than Jupiter. Hence, if discs have low viscosities we may not need to predict planets on the order of Jupiter's mass to explain potential vortices in observational results.

\subsection{Larger aspect ratio}\label{SubSec:AspRat}

So far we have only discussed the formation of vortices in a disc with $h = 0.05$. It is expected that as the disc aspect ratio changes, so will the planet growth timescale required for a planet to be formed without creating vortices, for a given viscosity and planet mass. We have investigated this behaviour, increasing the aspect ratio of the disc from $h = 0.05$ to $h = 0.06$. We only investigate one different aspect ratio and only for a small number of setups. Adding this additional parameter to our investigation into $q_f$ and $\nu$ would be infeasible due to increased computational time required. However, we can make a physically motivated guess at how the results vary by changing $h$ and then perform a number of simulations to check this. 

\begin{figure}
	\includegraphics[width=\columnwidth]{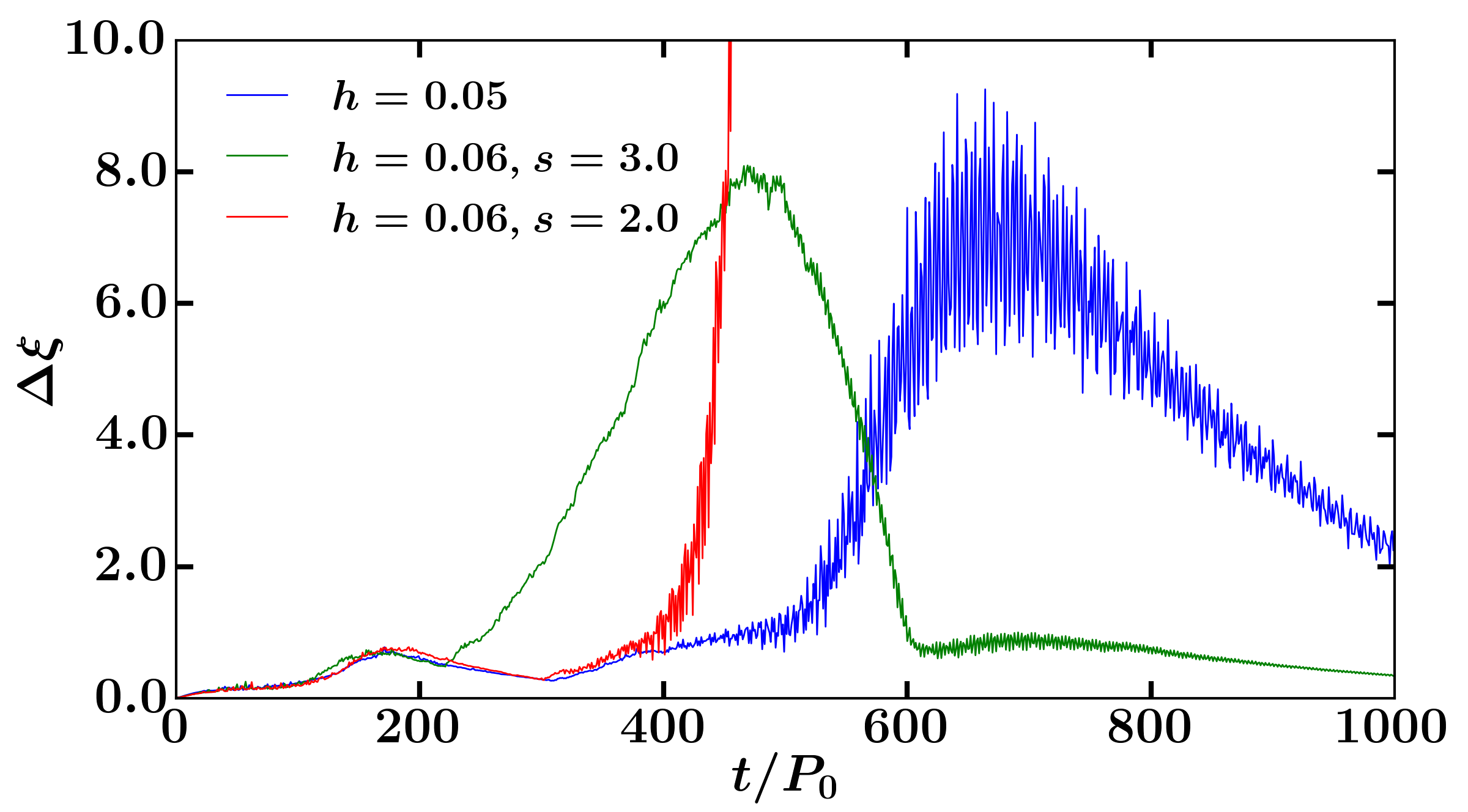}
	\vspace*{-5mm}
    \caption{Orbit-to-orbit maximum potential vorticity difference for the parameters discussed in Section \ref{SubSec:AspRat}, where $s$ represents the power of $h$ in the parameter setting relation. Here we can see that it is possible to get similar results despite increasing the aspect ratio by holding $K$ constant, however depending on how the parameters are selected the system may be forced more towards stability or vortex formation.} 
    \label{Fig:asprat}
\end{figure}

We know that the gap depth is set by the $K$ parameter \citep{Kanagawa2015},

\begin{equation}
	K = q_f^2\left(\frac{R_0}{h}\right)^5\alpha^{-1},
\end{equation}
where $\nu = \alpha c_s H$. Hence, we may expect that holding the $K$ parameter constant while increasing the aspect ratio gives similar results. However, a constant $K$ at a given $h$ is degenerate in $q_f$ and $\nu$ and therefore we need a method of selecting values for these parameters. To this end we use the relation $q_f/h^{3}$ to set $q_f$ at the increased aspect ratio. This relation governs the non-linearity of the flow in the inviscid case \citep{Korycansky&Papaloizou1996}. Using this we keep the non-linearity of the flow constant, while now having specific values for the parameters $q_f$ and $\nu$ at our selected value of $h = 0.06$. With these parameters we find a similar result to the $h = 0.05$ case, namely an unstable run remains unstable and a stable run remains stable, however using these parameters does seem to slightly improve the overall stability of the system compared to the $h=0.05$ case. Therefore, we investigate using the relation $q_f/h^{2}$ to select $q_f$ and $\nu$ instead, still with $h=0.06$. Using these parameters we found had the opposite effect, driving the system more towards instability and vortex formation compared to the respective $h=0.05$ case. These results can be seen in Figure \ref{Fig:asprat} which shows the orbit-to-orbit potential vorticity difference that we use as a tracer for vortex formation for both $h = 0.06$ cases and the $h=0.05$ case. The $h=0.05$ setup has $\nu = 9.0\times 10^{-6}$ and $q_f = 9.0\times 10^{-4}$, while all setups have $t_G = 600$. Clearly it can be seen that the $q_f/h^{2}$ case is driven more towards vortex formation, while the $q_f/h^{3}$ case is closer to reproducing the same results as $h=0.05$. The earlier increase in potential vorticity difference occurs as the gap opens faster in the higher planet mass cases. From these results we can see that using the relation $q_f/h^{3}$ gives reasonably consistent results with $h=0.05$, despite the increase to $h=0.06$, whereas $q_f/h^{2}$ is less consistent. However, we do not know if this result is applicable in every setup.

\section{Comparison with Observations}\label{Sec:Comp}
In the previous section we have found an approximate relation between the mass of a planet and the viscosity of the disc that predicts the growth timescale of the planet required in order for the planet to form without creating vortices within the disc. This is a result which allows us to make comparisons with observational data. Specifically we can use observational evidence for both the presence and absence of vortices in protoplanetary discs that show structure indicative of embedded planets. From these we can make tentative statements about the parameters of the discs and planets in question. 

An important consideration in our analysis here is the choice of growth timescale for planets within the discs. Whereas planet masses and disc viscosities are often predicted parameters from observational results, which we can use in our comparison, planet growth timescale is not often considered. As a result, for the comparisons in this section we use the age of the system as a limit on the planet growth timescale. For a planet in a disc that has reached maximum mass, the growth timescale of the planet cannot exceed the lifetime of the disc. Hence the growth timescales considered here represent the maximum possible growth timescale for a planet in that disc. As we use ages considerably longer than the growth timescales fit by Equation \ref{Eq:Pred_Fit}, we have to assume that our fits are still applicable in this regime.

Additionally, we have generally endeavoured to make statements using the units of mass presented in the papers to which we are comparing. However, for ease of comparison we also provide the conversion to mass ratio $q$ alongside each. With regards to viscosity, many of the papers to which we are comparing use a turbulent $\alpha$ viscosity \citep{ShakuraSunyaev1973}. We provide the conversion the the kinematic viscosity given the disc parameters used in our setup, at the location of the planet $R = R_0$.

\subsection{Axisymmetric discs}

\subsubsection{HL Tau}

\begin{figure}
	\includegraphics[width=\columnwidth]{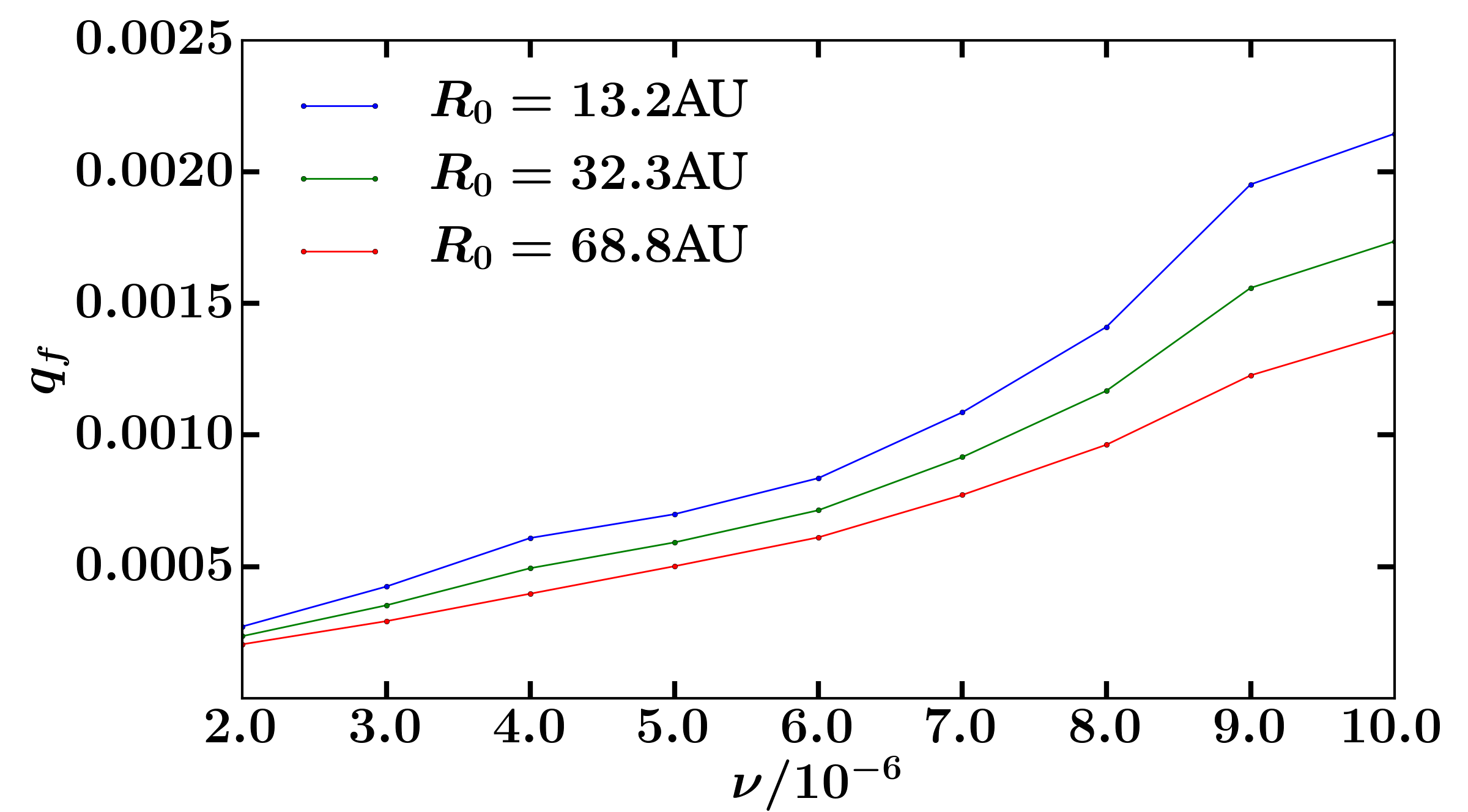}
	\vspace*{-5mm}
    \caption{Maximum masses of planets that can be formed over a timescale $t_{\textrm{G}}$ in a disc of a given viscosity. $t_{\textrm{G}}$ is set by the number of orbits able to be completed by a planet orbiting at a radius $R$ over the lifetime of the disc. This means that masses above this line will not form at this radius without creating vortices, while masses below will, according to our result in Equation \ref{Eq:Pred_Fit}. In this case the three radius' chosen correspond to the predicted locations of planets in the disc HL Tau \citep{Dipierro2015}.} 
    \label{Fig:HL_Tau}
\end{figure}

One of the most well documented protoplanetary discs of the last few years is HL Tau \citep{ALMApartnership2015}, for good reason as it is one of the highest resolution images of a protoplanetary disc we currently have, in addition to showing very interesting structure. The origin of this structure is still debated, as there are a number of different proposed factors that can explain the gap structure observed, besides planets. Nevertheless there are some factors that support the hypothesis that planets have carved these gaps, as the locations of these dark bands are located close to the resonances between planets orbiting at these radial locations \citep{Wolf2002,ALMApartnership2015}. \cite{Dipierro2015} performed smoothed particle hydrodynamics simulations to estimate the masses of planets that could potentially explain this structure. Their results imply that the observed structure could be the result of three planets of masses $q = 2.0\times 10^{-4}$, $2.7\times 10^{-4}$ and $5.5\times 10^{-4}$ located at $13.2$, $32.3$ and $68.8\textrm{AU}$ respectively. This prediction is very interesting to us, as crucially the HL Tau system shows no sign of vortices. Using Equation \ref{Eq:Pred_Fit} and assuming $t_{\textrm{G}} = 1\textrm{Myr}$ (an upper limit on the age of the HL Tau system \citep{Testi2015}), we can determine if our predictions regarding vortex formation agree with their planet masses. The results of this can be seen in Figure \ref{Fig:HL_Tau}, which shows the solution to Equation \ref{Eq:Pred_Fit} for a range of viscosities. \cite{Dipierro2015} use a turbulent viscosity \citep{ShakuraSunyaev1973} of $\alpha \approx 0.005$ ($\nu = 1.25\times 10^{-5}$), however more recent results from dust settling models imply that for HL Tau $\alpha \approx 10^{-4}$ ($\nu = 2.5\times 10^{-7}$) is an upper limit \citep{Pinte2016}. \cite{Dipierro2015} use a flared disc with aspect ratio $h_{\textrm{in}} = 0.04$ and $h_{\textrm{out}} = 0.08$.

From Figure \ref{Fig:HL_Tau} we can see that our results are consistent with this prediction for the very large viscosity cases. For $\nu > 8\times 10^{-6}$ we predict it is possible to form planets up to $\approx 1M_{J}$ ($q=1.0\times 10^{-3}$) at all of these radii with no vortices created. Using $\nu = 2.5\times 10^{-7}$ it appears that the planets predicted by \cite{Dipierro2015} are unable to form without creating vortices. Our results predict that planets with masses of roughly $q = 2.5 \times 10^{-5}$ are the largest that could form here without creating vortices. Using the viscosity from \cite{Dipierro2015} however it is very clear from our results that their planets could form without creating vortices.

\cite{Jin2016} make predictions in a lower viscosity regime than \cite{Dipierro2015}. They imply that the observed structure can be explained by planets of masses $0.35M_{J}$, $0.17M_{J}$ and $0.26M_{J}$ ($q = 3.5, 1.7$ and $2.6\times 10^{-4}$ respectively) at approximately the same radii as above. \cite{Jin2016} use a disc viscosity of $\alpha = 10^{-3}$ ($\nu = 2.5\times 10^{-6}$) which lies in between the above viscosities. Fortunately, this viscosity also lies within our explored parameter space. From Figure \ref{Fig:HL_Tau} we can see that at this viscosity we can expect planets of $q \approx 2.5-3.5\times 10^{-4}$ to form without creating vortices. Therefore, the planets of \cite{Jin2016} are roughly the maximum mass planets our results predict could be present in the HL Tau disc at this viscosity. Hence, our results show consistency with their predictions.

\subsubsection{TW Hydrae}

Another well documented disc that shows similar features to HL Tau is TW Hydrae, also the result of high resolution ALMA observations \citep{Andrews2016}. This disc also shows axisymmetric dark radial rings that could potentially be the result of gaps formed by planet disc interactions. An interesting difference between TW Hydrae and HL Tau is that while the latter is a young disc, TW Hydrae is particularly old, with an age of $3-15\textrm{Myr}$ \citep{Sokal2018}. Another difference that separates these discs is the dark bands themselves. In TW Hydrae these dark bands are significantly brighter (compared to the surrounding disc) and narrower than those in HL Tau. The masses of planets that could potentially explain these gaps has been investigated by \cite{vanBoekel2017}, using the results of \cite{Duffell2015} and was found to be in the range of several earth masses. This corresponds to roughly $4-10M_{\textrm{E}}$, $10-20M_{\textrm{E}}$ and $25-50M_{\textrm{E}}$ ($q(1M_\textrm{E}) \approx 3.0\times 10^{-6}$) at $6.0$, $21.0$ and $85.0\textrm{AU}$ respectively, for viscosities between $\alpha = 1.0 - 4.0\times 10^{-4}$ ($\nu = 2.5\times 10^{-7} - 1.0\times 10^{-6}$). From our results we predict that the gaps at $6$ and $21\textrm{AU}$ are capable of being opened by these planets without forming vortices (except for the gap at $21\textrm{AU}$ in the lowest viscosity case). However, our results imply that the larger mass planet at $85\textrm{AU}$ could not form there without creating vortices. These predictions are based on extrapolation, as we have not investigated at viscosities this low. \cite{Tsukagoshi2016} also imply that the most prominent, central gap could be caused by an approximately $1.5$ Neptune mass planet ($q = 7.72\times10^{-5}$) in an $h = 0.05$, $\alpha = 10^{-3}$ ($\nu = 2.5\times 10^{-6}$) disc. As this planet is low mass and TW Hydrae is an older star, our results predict that this planet could form without causing the creation of vortices. Hence we display some consistency here with previous studies.

\subsubsection{Taurus star forming region}

\begin{figure}
	\includegraphics[width=\columnwidth]{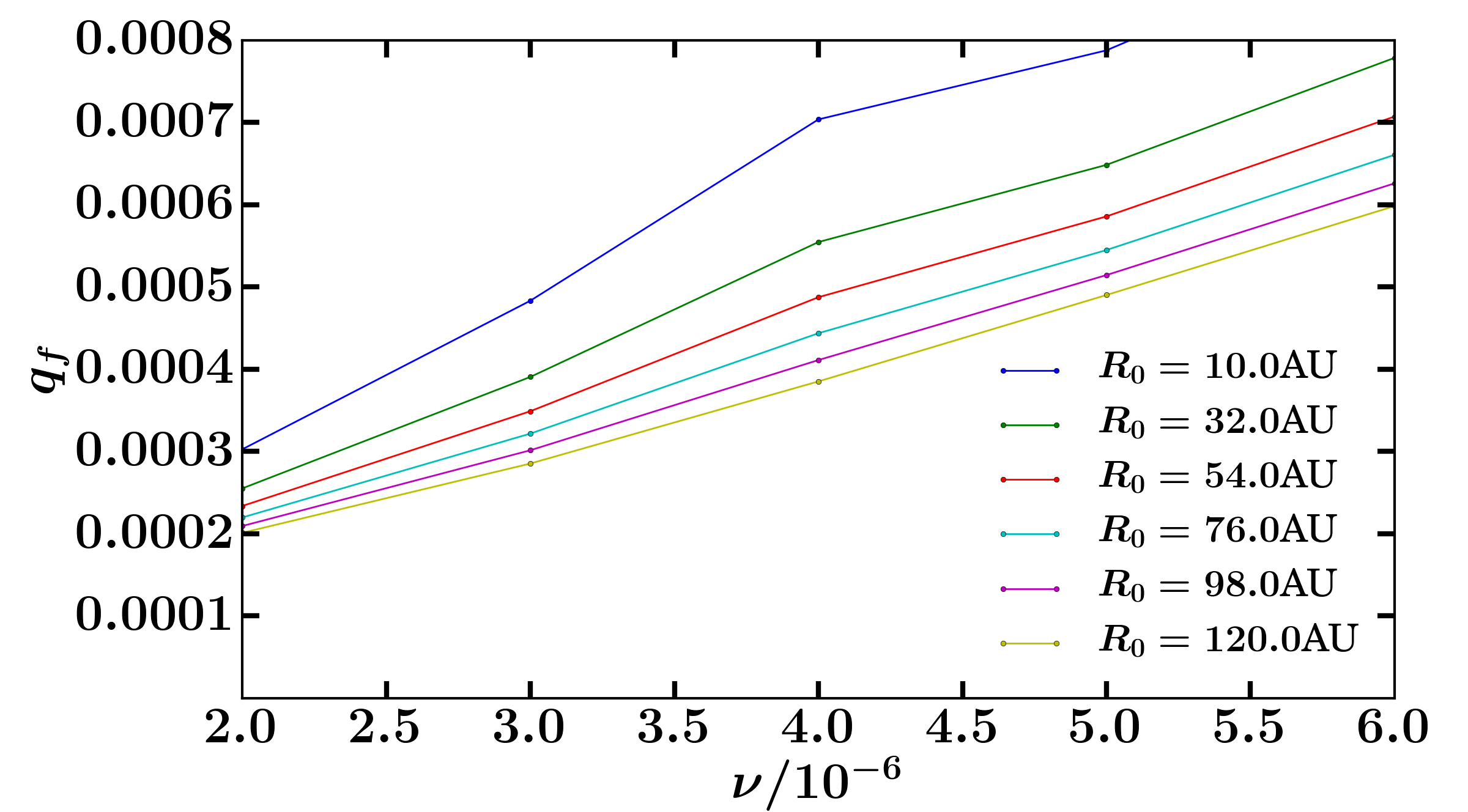}
	\vspace*{-5mm}
    \caption{Maximum masses that can be formed over a timescale $t_{\textrm{G}}$ in a disc of a given viscosity, similarly to Figure \ref{Fig:HL_Tau}. In this case the radius' chosen correspond to the predicted locations of planets in the sample of $12$ discs analysed by \citep{Long2018}. Specifically this plot focusses on the region for which planets of masses in the range $0.1-0.5M_J$ ($q = 1.0\times 10^{-4} - 5.0\times 10^{-4}$) can form.}
    \label{Fig:Sample_12}
\end{figure}

More recently, \cite{Long2018} performed an analysis on $32$ discs in the Taurus star forming region using ALMA. From this sample $12$ discs containing axisymmetric structure were identified. This structure takes the form of dark band bright ring pairs, emission bumps and cavities. Overall 19 gap ring pairs were identified, indicating that a number of these systems contain multiple gaps. These gaps range in location from $R = 10 - 120AU$ with no preferred distance. The majority of these gaps are narrow, but the weak correlation between gap location and gap width potentially implies formation via planet-disc interactions. In addition a significant number of these gaps can not be explained by condensation fronts. \cite{Long2018} perform an analysis on the width of these gaps in order to determine possible masses of planets that could form them. Assuming the width of the gaps corresponds to $4R_{\textrm{Hill}}$ \citep{Dodson-Robinson&Salyk2011} they estimate the masses of the planets to be in the $0.1-0.5M_J$ ($q = 1.0\times 10^{-4} - 5.0\times 10^{-4}$) range, however they stress that these masses have large uncertainties and that the gap widths may be as large as $7-10R_{\textrm{Hill}}$ \citep{Pinilla2012}. Alternatively they compare the distance between the minimum of the gaps and the maximum of the rings to the results of hydrodynamic simulations \citep{Rosotti2016} with an $\alpha = 10^{-4}$ ($\nu = 2.5\times 10^{-7}$), resulting in a predicted planet mass of $0.05M_J$ ($q = 5.0\times10^{-5}$) and an $\alpha = 10^{-2}$ ($\nu = 2.5\times 10^{-5}$), resulting in a predicted planet mass of $0.3M_J$ ($q = 3.0\times 10^{-4}$). Using Equation \ref{Eq:Pred_Fit} and an estimated disc lifetime of $2\textrm{Myr}$ (based on stellar age estimates for the spectral type of the target stars \citep{Baraffe2015,Feiden2016}) we can see that in the higher viscosity case planets of $0.3M_J$ ($q = 3.0\times 10^{-4}$) can form at any of these radii without creating vortices. The lower viscosity case is more difficult to predict as it is significantly lower than the viscosities we have explored, however for this planet mass it seems unlikely from our results that the predicted planet could form without creating vortices. Instead, we expect the maximum mass planets that could be found here to be roughly half the mass they predict. In addition, from our results planets of mass $0.5M_J$ ($q = 5.0\times 10^{-4}$) would require the disc viscosity to be $\nu \gtrapprox 3.0 - 5.0 \times 10^{-6}$ to form without creating vortices, depending on the radius at which they are located. Conversely we find planets as small as $0.1M_J$ ($q = 1.0\times 10^{-4}$) can even form at $\nu = 2.0\times 10^{-6}$ without creating vortices. This can be seen in Figure \ref{Fig:Sample_12}. Hence our results show that these gaps could be opened by planets of the masses predicted by \cite{Long2018} without forming vortices.

\subsubsection{DSHARP}

Recently results from DSHARP (Disc Substructures at High Angular Resolution Project) have been published \citep{DSHARPI2018}. This is one of the first large programs carried out using the ALMA telescope. In this project millimetre emission from a sample of $20$ protoplanetary discs was observed with intent to find and characterise substructures, such as dark gaps and bright rings. It was found that these substructures are present at almost any radius across the whole sample of discs. They are significantly more common than spiral arm structure, occurring in $3$ of the discs and crescent asymmetries, occurring in $2$ of the discs. Both of these additional structures coexist with the dark band substructures present across the sample. Similarly to \cite{Long2018} it is found that condensation fronts are not a reliable method of forming these potential gaps \citep{DSHARPII2018}. 

\cite{DSHARPVII2018} analyse $14$ discs from this sample, in which a total of $19$ gaps are present. Relations between gap width/depth and planet/disc properties are found using two dimensional hydrodynamic simulations including dust particles for a range of disc aspect ratios, turbulent viscosities and planet masses appropriate to the sample. These relations are then used to infer the planet mass from the intensity profiles of \cite{DSHARPII2018}. However, only $12$ of these gaps are wide enough to be fit using their fitting formula, whereas the remainder are fitted by eye. Hence, we will focus our comparison here on the more robustly fit sample of $12$. 

The predicted planet masses listed in \cite{DSHARPVII2018} clearly highlight the degeneracy we are attempting to address. Not only does the planet mass differ depending on which turbulent viscosity is assumed but also the assumed dust size distribution. This results in between $3$ and $9$ possible planet masses that can explain each gap. While our results can do nothing to address the degeneracy arising from the assumed dust size distribution, we can make some predictions based on the turbulent viscosity. The turbulent viscosities assumed in their analysis are $\alpha = 10^{-4}, 10^{-3}$ and $10^{-2}$ ($\nu = 2.5\times 10^{-7}, \times 10^{-6}$ and $\times 10^{-5}$ respectively). For each of these viscosities the predicted masses range from $0.01-1.06M_{J}$, $0.03-2.16M_{J}$ and $0.06-4.41M_{J}$ ($q(1M_J) = 1.0\times 10^{-3}$) respectively in the axisymmetric discs. The aspect ratios of their discs range from $h = 0.04 - 0.09$. We perform a general overall comparison using a disc lifetime of $1\textrm{Myr}$ \citep{DSHARPI2018}, $h = 0.05$ and $R = 10 - 100\textrm{AU}$ as the locations of the gaps for our comparison. Using this our results show that in the lowest viscosity case, planets of up to $0.025M_{J}$ ($q = 2.5\times 10^{-5}$) can form without creating vortices and in the middle viscosity case planets of up to $0.36M_{J}$ ($q = 3.6\times 10^{-4}$) can form without creating vortices. In this case the lowest viscosity prediction requires some extrapolation to make as this viscosity lies outside the domain of our results. Similarly, the highest viscosity case is significantly outside the domain of our results, such that it is difficult to make meaningful extrapolations towards. At the edge of our results domain, at $\nu = 10^{-5}$, planets of up to $2.0M_{J}$ ($q = 2.0\times 10^{-3}$) can form without creating vortices. These are the maximum masses in each case, which corresponds to the smallest radial location of the gap and increasing the radial location will reduce the mass of a planet that can form. In general for the lower viscosities this mass reduction is very small. Hence, our results provide a limit which can rule out the low viscosity, high planet mass cases. 

Finally we must address the point that two of the discs in this sample (HD143006 and HD163296) show asymmetries and hence require a slightly different analysis. In HD143006 the asymmetry lies beyond the continuum disc edge and such is not a direct analogue to our gap edge vortices. However, this disc is predicted to contain a planet of mass $\approx 2-40M_{J}$ ($q = 2-40\times 10^{-3}$) dependent on the degeneracies listed above, which would be impossible to form without creating vortices according to our results. The asymmetry in HD163296 is more consistent with those we see in our simulations, occurring at the $R=48\textrm{AU}$ gap in this disc. Using this gap location and an age of $10\textrm{Myr}$ \citep{DSHARPI2018} we find that all of the predicted masses $0.54-4.45M_{J}$ ($q = 0.54-4.45\times 10^{-3}$) across the above degeneracies will not be able to form without creating vortices. The only potential exception to this is in the high viscosity case, which suffers from the same extrapolation problem as in the comparison above. At $\nu = 10^{-5}$ it is already possible for their smallest predicted planet to form while retaining axisymmetry, for their highest viscosity. This parameter space will very likely become larger as the viscosity increases. Nevertheless, the overall agreement with this disc is very good.

Another very interesting result from the analysis of \cite{DSHARPVII2018} is for the specific case of the disc AS209 which contains axisymmetric gaps at $9$, $24$, $35$, $62$, $90$, $105$ and $137\textrm{AU}$. They find that a planet of mass $q = 1.0\times 10^{-4}$ at $R = 100\textrm{AU}$ in a disc with $\alpha = 10^{-5}$ ($\nu = 2.5\times 10^{-8}$) and $h = 0.05$ can very well explain both position and amplitude of the $24$, $35$, $62$, $90$ and $105\textrm{AU}$ gaps. The age of this disc is $1\textrm{Myr}$ \citep{DSHARPI2018}. Again we must extrapolate as the viscosity is so small, but unfortunately from our results it is very clear that a planet of this mass would not be able to form in these conditions without also creating vortices. Interestingly, despite reproducing well the location and amplitude their results also show some asymmetry in the disc substructure. This is overcome by using a radially varying turbulent viscosity, which also provides a better fit to the observations. Hence, this is something that may need to be considered.

\subsection{Non-axisymmetric discs}

\subsubsection{Oph IRS 48}

There are many observational examples of discs containing vortices that could potentially be the result of planet disc interactions \citep[e.g.][]{vanderMarel2013,Isella2013,Fukagawa2013,Perez2014,Marino2015,Kraus2017,Dong2018,Cazzoletti2018}. These are also particularly interesting cases to us as, similarly to the HL Tau and TW Hydrae cases, allow us to make tentative predictions about the planets in these systems. \cite{vanderMarel2013} investigate the disc surrounding the young A type star Oph IRS 48, in which a dust trap is located between $45$ and $80\textrm{AU}$ at the edge of an inner cavity. They suggest a planet of mass $\approx 10M_{\textrm{J}}$ ($q = 1.0\times 10^{-2}$) located at $20\textrm{AU}$ could explain this and that the dust trap at the edge of the cavity could be a vortex arising from the excitation of the Rossby wave instability. In their simulations, they use an $\alpha = 10^{-4}$ ($\nu = 2.5\times 10^{-7}$) and an $h = 0.05$ at the location of the planet. This aspect ratio is consistent with our investigation, however their disc is flaring. From Equation \ref{Eq:Pred_Fit} we can see that attempting to form such a massive planet over standard young disc lifetimes ($1-5\textrm{Myr}$) will always result in vortices.

\subsubsection{HD135344B}

The disc surrounding the Herbig Ae/Be star HD135344B \citep{Perez2014, Cazzoletti2018} shows interesting phenomena. This disc contains a cavity with a radius of $\approx 50\textrm{AU}$ at the edge of which is a bright ring of material. At $80\textrm{AU}$ there is a crescent asymmetry. The disc also contains two bright spiral arms extending out to $\approx 75\textrm{AU}$. This disc is also particularly interesting as it is currently the only observational result we have that shows the features identified by \cite{Hammer2019}, pertaining to an elongated vortex forming as a result of a planet growing in the disc. \cite{Fung&Dong2015,Dong&Fung2017,Cazzoletti2018} show that these phenomena could possibly be explained by a planet of $5.0M_{\textrm{J}}$ ($q = 5.0\times 10^{-3}$) situated between the bright ring and the asymmetry. For standard disc lifetimes $1-10\textrm{Myr}$ and viscosity $\nu$ on the order of a few $10^{-6}$ we can confirm from Equation \ref{Eq:Pred_Fit} that forming such a massive planet at this location will create vortices. Also particularly interestingly \cite{Cazzoletti2018} show that, by using the results of \cite{Rosotti2016,Facchini2018}, the gap between these features can be explained by a $0.2M_{\textrm{J}}$ ($q = 2.0\times 10^{-4}$) planet at $68\textrm{AU}$ with a viscosity of $\alpha = 10^{-3}$ ($\nu = 2.5\times 10^{-6}$). In this situation the planet is not massive enough to explain the spiral arms, hence is not a result that explains all of the features in the disc. Our results show the formation of this planet should not create vortices for ages $1-10\textrm{Myr}$. This is interesting, as while it has been predicted that a planet of this mass could form the gap we see, both our results and the analysis by \cite{Cazzoletti2018} predict that this planet could not form the additional features present in this disc. This is an example of using our results to rule out a planet of a certain mass as an explanation for a gap in a disc, despite this planet being a reasonable candidate when a large number of disc parameters are unknowns.

\subsubsection{MWC758}

Another system of interest surrounds the Herbig Ae/Be star MWC758 \citep{Marino2015, Dong2018}. This star is $3.7 \pm 2\textrm{Myr}$ old and has a cavity of radius $50\textrm{AU}$ in which there is the possibility of a low mass companion. The outer disc extends to $\approx 100\textrm{AU}$ with two clumps at $50$ and $85\textrm{AU}$. \cite{Dong2018} postulate that a $5-10M_{\textrm{J}}$ ($q = 5-10\times 10^{-3}$) planet at $100\textrm{AU}$ could be the cause of this \citep{Dong2015}. Once again for a reasonable range of disc viscosities, using the age of the star as an upper limit we find from Equation \ref{Eq:Pred_Fit} that attempting to form a planet this massive at this radius will create vortices.

\subsection{Concluding Remarks}

The majority of our investigation has been undertaken assuming a constant disc aspect ratio $h=0.05$, however as discussed in Section \ref{SubSec:AspRat} we have investigated the effect of varying this parameter. As the results discussed here are observational, it is very difficult to know the aspect ratios of the discs in question. As a result predictions made regarding masses of gap opening planets will often assume the value of $h$. Our investigation into the aspect ratio shows that a thicker disc will tend to be more stable, for the same viscosity and planet mass. Therefore, should the discs in these observations have a larger aspect ratio than has been assumed (or the assumed aspect ratio is greater than $h=0.05$) the planet masses and viscosities could be larger before vortices are created. This also means that if we predict a planet of a given mass can form in a disc without causing vortices to appear, then that same planet can still form without creating vortices if the disc's aspect ratio was larger than we assumed.

The findings from our results and the predictions regarding the above systems is twofold. Firstly it is apparent that given our relation and for reasonable ranges of disc viscosity and age, several $M_{\textrm{J}}$ planets will always create vortices in the disc during their formation. Hence in these discs we have observed, if we expect to see a planet of multiple $M_{\textrm{J}}$ in these discs we should also expect to find asymmetries. Secondly, our results show that vortices are formed for planets significantly smaller than multiple $M_{\textrm{J}}$ for low viscosities. This implies that, in systems observationally showing signs of vortices, the planets may not be as giant as we previously expected. Hence, vortices in discs may not be a tracer of multiple $M_{\textrm{J}}$ mass planets, instead they could show the presence of planets with masses in the sub-Saturn range, down to even only a few Neptune masses. In addition, the absence of vortices in low viscosity discs implies that if any planets exist in the disc, they must be very low mass.

\section{Discussion}\label{Sec:Disc}

Our results and the above conclusion are still very tentative. We have made a number of assumptions in our simulations that could have some impact on the validity of our main result, in addition to the aforementioned errors. We will justify and discuss these assumptions here.

We make no statements regarding how these planets could form at these locations within the disc. We use a very simplified method of growing the planet mass over a timescale, discussed in Section \ref{SubSec:GrowP}, that does not take into account any particular formation mechanism. There are two main theories as to how giant planets form in protoplanetary discs, the core accretion and gravitational instability theories \citep{Pollack1996,Boss1997,Helled2014}, both of which contain their own problems that can be difficult to solve. In neither case do we expect the growth of the planet to follow Equation \ref{Eq:GrowP}. However, we expect that it is the growth timescale that is the most important parameter in determining whether the gap edge becomes unstable and any method of planet formation will have a corresponding growth timescale.

We also assume our planet does not migrate in the disc. Planet migration would clearly impact the planet formation process, as it is more difficult to form giant planets at large radial distances in the disc. As we expect planets to migrate inwards this means that it will be easier to form giant planets as the migrating planet approaches the warmer, more dense regions of the disc. Again, similarly to planet formation there are different regimes of planet migration through which a planet will progress as its mass increases. These are Type I migration for low mass planets \citep{Goldreich&Tremaine1980} and Type II migration for more massive, gap opening planets \citep{LinPapaloizou1986}. As a planet grows it will initially migrate at high speed via Type I migration until it begins to open a gap in the disc and transfers to Type II migration, at which this speed will be reduced \citep{Ward1997}. Despite this reduction in migration rate the predicted Type II migration rate is still shorter than the lifetime of the disc, implying that most planets undergoing Type II migration will be absorbed by the central star \citep{Hasegawa&Ida2013}. This is problematic for planet formation models, as observational evidence dictates that gas giant planets are more common at orbital distances $R > 1\textrm{AU}$ \citep{Mayor2011,Cassan2012,Fressin2013,Santerne2016}. Which could not occur if this predicted timescale is correct. Current models predict that giant planets must form at $R > 20\textrm{AU}$ to survive this migration \citep{Coleman&Nelson2014}. Hence it is expected that something is limiting the rate of Type II migration \citep{Nelson2000}. Due to its complex nature, we neglect planet migration in this initial study and leave accounting for it to future work. 

The impact of the Rossby wave instability on migrating planets is still largely unexplored, however, there has been some recent investigation into its effects \citep{McNally2019}. As the planet's wake migrates with the planet, the effect of the shock will be more spread out over the disc, possibly leading to a gap edge that is less steep. Hence, it is possible that the migration of the planet may act to suppress the Rossby wave instability. Additionally, the relation between gap opening and planet migration is not well understood. It has been shown that for a migrating planet to open a gap, the gap formation timescale must be shorter than the migration timescale, otherwise a gap will not be formed \citep{Malik2015}. This can be difficult considering the rapid rate of planet migration. Also, growing a planet in the disc will cause the gap formation timescale to be longer than if the planet was initialised at its final mass. If the migrating planet cannot open a gap, then the Rossby wave instability will not be excited as no steep gap edge is formed. Nevertheless, it has been shown that migrating giant planets can still open gaps in the disc material \citep{Crida&Bitsch2017}.

In addition to our simplistic approach to planet growth and migration, the systems we simulate are significantly less complex than those we are comparing our results to. We use a constant viscosity across our disc and hold our planets on a circular orbit with zero eccentricity. Most importantly we only consider a single planet within the disc, whereas observationally we often see tracers of multiple planets, such as the multiple dark bands in both HL Tau and TW Hydrae. However, in discs such as HL Tau the dark bands are far apart, so that if they were gaps formed by planets we can safely treat them as seperate entities. Also, the vortices formed as a result of instability excitation in our simulations always occurs on the outer edge of the gap formed by the planet. This is often the case in observational results, but not always as we can also see vortices formed at the edge of cavities and also interior to the planet. This should have little implication for comparison with our results, but shows that there are differences between our simulations and the systems we observe.

We perform our simulations using a simplified, locally isothermal equation of state. In further work we intend to extend this investigation to a non-isothermal equation of state, using a cooling timescale. However, when considering non-isothermal discs, the cooling timescale is a parameter that is difficult to estimate as it is dependent on disc opacity. The cooling timescale can be chosen to ensure initial equilibrium between (viscous) heating and cooling \citep[e.g.][]{Pierens&Lin2018}. More realistic cooling prescriptions can be constructed under the same assumption \citep[e.g.][]{Faure2015}.

Another factor that may impact the validity of our comparisons with observational results is that we simulate only the gas in the disc, while observationally we can only view the emission from the dust in the disc. This means that unless the dust and gas behave very similarly there may be some difference between our predictions and observations. It is already known that it is easier to form gaps in dust than in the gas \citep{Paardekooper&Mellema2004}, therefore it is possible that gaps we see from observational results may still contain gas.

We make no predictions concerning the lifetime of the vortices arising as the result of instability excitation in our simulations. We find that some vortices formed during planet growth will mostly disperse before the simulation has elapsed, but in other cases vortices will still be present when the simulation has been finished. We note that the lifetime of a vortex is difficult to predict. There are a number of factors that can reduce the lifetime of vortices, such as the viscosity \citep{DeValBorro2007,Ataiee2013,Fu2014,Regaly2017}, dust feedback \citep{Johansen2004,Inaba&Barge2006,Lyra2009,Fu2014_2}, disc self gravity \citep{Regaly&Vorobyov2017,Pierens&Lin2018} and the streaming instability \citep{Raettig2015}. The lifetimes of vortices formed by giant planets has been investigated and were found to be strongly dependent on planet mass and viscosity, capable of sustaining vortices up to $10^{4}$ orbits in some cases \citep{Fu2014}. It has also been shown that dust build up in vortices can act to speed up their dispersion \citep{Railton&Papaloizou2014}. Vortices in protoplanetary discs are very efficient at trapping dust \citep{Barge&Sommeria1995}, however the dust feedback can destroy the vortex if the dust density becomes too large. This occurs due to the excitement of a dynamical instability which destroys the potential vorticity minimum, a prerequisite for sustaining the vortex. Hence when this is removed the vortex quickly dissipates \citep{Fu2014_2}. It has also been shown that a dust vortex will live longer than a gas vortex, hence after the gas vortex dissipates we expect to still see asymmetries in the dust. The lifetime of asymmetries in the dust can be four times longer than those in the gas \citep{Fu2014_2}. As our simulations only consider the gas in the disc, it is possible that the vortices in the dust would be observable for a longer period. However, it is also possible that the vortices would dissipate faster if we considered their tendency to trap dust. If a brightness asymmetry exists in an observation, we know there is the potential for a vortex to exist and can use this to constrain the mass of any potential planets in the disc. If we do not see a potential vortex, then we can use their absence to constrain the mass of any potential planets in the disc. However, we must be aware that it is possible for a vortex to have existed earlier in the disc's lifetime and has since dissipated. This is an unfortunate consequence of only being able to observe a disc at a single point during its evolution. Additionally, dust enhancement of vortices can make these vortices more visible in the dust than in the gas. As we have discussed, observational results show the dust in the disc, whereas our simulations represent the gas in the disc. Therefore, it is possible that weak vortices in our simulations could correspond to more visible vortices after accounting for dust enhancement. Hence, we have been particularly careful analysing marginal cases with weak vortices in establishing the planet growth timescale at which vortices disappear. 

Finally, we do not include any effects of disc self gravity in our model. However, it is known that disc self gravity can have major effects on vortex formation and lifetime \citep{Regaly&Vorobyov2017}. It is also known that the indirect potential caused by a vortex can aid the onset of the Rossby wave instability \citep{Regaly&Vorobyov2017a}. Hence, our results are not applicable to massive discs for which self gravity becomes important, but this would be an interesting direction to take this study.
 
\section{Conclusions}\label{Sec:Conc}
In this paper we have derived a relation between the viscosity of a disc and the mass of planets forming in this disc as a function of the growth timescale of the planet. We intend this relation to help break the degeneracy in planet masses and viscosities that can adequately explain observed gaps in protoplanetary discs. If both of these parameters are unknown then almost any gap can have a planetary explanation. Using this relation we can make predictions regarding the mass of planets in protoplanetary discs in which we observationally detect the presence of absence of vortices, assuming standard disc ages and viscosities. We find that our results are consistent with predictions made regarding a variety of observations, both for discs containing and devoid of vortices, including well documented discs such as HL Tau \citep{ALMApartnership2015} and TW Hydrae \citep{Andrews2016}. However, we must stress that our relation is very tentative and is by no means a definitive prediction, due to both the inherent inaccuracies in the method it was derived and the assumptions made in our simulations.

Furthermore, using our relation we can make the prediction that it is not necessary for planets to have masses on the order of one Jupiter mass to form vortices within low viscosity discs. While massive planets do indeed form these vortices, the onset of them occurs for planets that are significantly less massive, for acceptable disc viscosities and ages. We find that planets of sub-Saturn mass can form vortices, down to a few Neptune masses. Whether or not planets of these low masses can explain the dark bands we see in many of these observations is a different question. We are in no way precluding giant Jupiter mass planets from being the cause of these, however this prediction must be something to keep in mind, especially as we delve deeper into attempting to directly image these planets within their gaps. We also hope that our results are a step on the way to breaking the degeneracy in planetary explanations for observational gaps, by imposing limits on the mass of the planet and/or the viscosity of the disc based on the presence or absence of vortices in the disc.

\section*{Acknowledgements}
Sijme-Jan Paardekooper is supported by a Royal Society University Research Fellowship. We thank the reviewer Zsolt Reg\'aly, for his insightful comments.



\bibliographystyle{mnras}
\bibliography{mnrasRef}

\bsp	
\label{lastpage}
\end{document}